\documentclass[aps, prl,twocolumn]{revtex4-1}

\usepackage[colorlinks=true,citecolor=red]{hyperref}
\usepackage{verbatim}
\usepackage{stmaryrd}

\usepackage{tikz}
\usetikzlibrary{decorations.pathreplacing}
\pgfrealjobname{polar-prl}

\usepackage{xspace}
\usepackage{amsmath}
\usepackage{amssymb}
\usepackage{bbm}
\usepackage{braket}
\usepackage{xfrac}

\usepackage[resetlabels]{multibib}
\newcites{supp}{Supp}

\newcommand{\arikan}{Ar\i{}kan{}}
\newcommand{\prlsection}[1]{\emph{#1}.---}
\def\cnot{\textsc{cnot}\xspace}
\renewcommand{\rho}{\varrho}

\def\mcA{\ensuremath{\mathcal{A}}\xspace}
\def\mcE{\ensuremath{\mathcal{E}}\xspace}
\def\mcP{\ensuremath{\mathcal{P}}\xspace}
\def\mcQ{\ensuremath{\mathcal{Q}}\xspace}
\def\mcM{\ensuremath{\mathcal{M}}\xspace}
\newcommand{\wt}[1]{\widetilde{#1}}

\newcommand{\sketbra}[2]{{\ensuremath{\lvert #1\rangle\!\langle #2\rvert}}}
\newcommand{\lketbra}[2]{{\ensuremath{\left\lvert #1\middle\rangle\!\middle\langle #2\right\rvert}}}
\newcommand{\ketbra}[2]{\mathchoice{\lketbra{#1}{#2}}{\sketbra{#1}{#2}}{\sketbra{#1}{#2}}{\sketbra{#1}{#2}}}

\begin{document}

\author{Joseph M.\ Renes}
\author{Fr\'ed\'eric Dupuis}
\author{Renato Renner}
\affiliation{Institut f\"ur Theoretische Physik, ETH Zurich, CH-8093 Z\"urich, Switzerland}

\title{Efficient Polar Coding of Quantum Information}
\begin{abstract}
  Polar coding, introduced 2008 by \arikan, is the first (very) efficiently encodable and decodable coding scheme whose information transmission rate provably achieves the Shannon bound for classical discrete memoryless channels in the asymptotic limit of large block sizes. Here we study the use of polar codes for the transmission of quantum information. Focusing on the case of qubit Pauli channels and qubit erasure channels, we use classical polar codes to construct a coding scheme which, using some pre-shared entanglement, asymptotically achieves a net transmission rate equal to the coherent information using efficient encoding and decoding operations and code construction.  Furthermore, for channels with sufficiently low noise level, we demonstrate that the rate of preshared entanglement required is zero.
\end{abstract}

\maketitle

One of the most exciting developments in classical information theory of the last decade, polar coding is a channel-adapted, block coding scheme which enjoys essentially all of the features one would like such schemes to have~\cite{arikan_channel_2009}. Polar codes enable the transmission of information over discrete, memoryless channels (DMCs) at rates up to the symmetric capacity of the channel (the Shannon limit  
assuming uniformly-distributed inputs to the channel)~\cite{sasoglu_polarization_2009}, and the capacity can be reached for arbitrary channel noise rates. Just as important, both the construction of polar codes as well as the encoding and decoding operations can be performed very efficiently, in  $O(n\log n)$ steps for $n$ the blocklength of the code~\cite{arikan_channel_2009,mori_performance_2009}. 

The main idea underlying the construction of polar codes is
\emph{channel polarization}: Out of $n$ identical DMCs one can create a
new set of $n$ logical channels via a suitable transformation such that each
logical channel is essentially either ``good''
(nearly noiseless) or ``bad'' (completely noisy). Messages can then be transmitted via the good channels, while the inputs to the bad channels   are fixed or ``frozen'' to values known to the decoder. 
For $n\rightarrow \infty$ the fraction of good channels
approaches the symmetric capacity of the original DMC, and thus the coding scheme achieves the symmetric capacity. 

Polar codes have attracted
considerable interest in the classical information theory community.
For instance see \cite{mori_performance_2009} for efficient constructions of polar
codes, \cite{korada_polar_2010} for bounds on their error probabilities, 
\cite{hussami_performance_2009,korada_polar_2009} for their use for source coding, and~\cite{mahdavifar_achieving_2011} for applications to the private communication over the wire-tap channel;~\cite{korada_polar_2009} also provides an excellent overview. 
Recently, Wilde and Guha showed that channel polarization extends to quantum channels when transmitting either classical~\cite{wilde_polar_2011-1} or quantum~\cite{wilde_polar_2011} information, but did not give an efficient decoding algorithm. 

In this Letter we show how classical polar codes can be adapted to the task of efficiently transmitting quantum information over noisy channels. Our construction is specifically formulated for qubit channels, and the resulting codes are CSS codes~\cite{calderbank_good_1996,steane_multiple-particle_1996}. For Pauli channels and the erasure channel, we show that quantum information can be transmitted at a rate given by the symmetric coherent information, the coherent information of the channel evaluated for Bell-state input, again using efficient encoding and decoding operations. Generically our construction  requires the use of preshared entanglement  between sender and receiver, though we demonstrate that in many cases the rate of preshared entanglement required is zero.

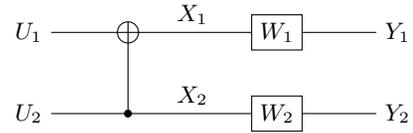
\begin{figure}[h]
\centering

\beginpgfgraphicnamed{chantrans}
\def\vgap{.9}
\def\hgap{1.1}
\def\hgapfudge{.25}
\begin{tikzpicture}[scale=1.2]
\tikzstyle{empty} = [inner sep=1pt,outer sep=1pt]
\tikzstyle{gate} = [fill=white, draw]
\tikzstyle{ctrl} = [fill,shape=circle,minimum size=3pt,inner sep=0pt,outer sep=0pt]
\tikzstyle{targ} = [draw,shape=circle,minimum size=8pt,inner sep=0pt,outer sep=0pt]

\node[anchor=east] at (0+\hgapfudge,0) (u2) {$U_2$};
\node[anchor=east] at (0+\hgapfudge,\vgap) (u1) {$U_1$};

\node[anchor=west] at (3.5*\hgap,0) (y2) {$Y_2$};
\node[anchor=west] at (3.5*\hgap,\vgap) (y1) {$Y_1$};

\node[anchor=south] at (1.65*\hgap,0) {$X_2$};
\node[anchor=south] at (1.65*\hgap,\vgap) {$X_1$};

\draw (u2) -- (y2);
\draw (u1) -- (y1);

\node[ctrl] at (\hgap,0) (ec) {};
\node[targ] at (\hgap,\vgap) (et) {};
\draw (ec) -- (et.north);

\node[gate] at (2.5*\hgap,0) (w2) {$W_2$};
\node[gate] at (2.5*\hgap,\vgap) (w1) {$W_1$};

\end{tikzpicture}
\endpgfgraphicnamed
\caption{\label{fig:channeltransform} The basic polar coding channel transformation. Two instances of a channel $W$ are transformed into two logical channels, one with higher information-carrying capacity than the original and the other lower. The worse channel takes $U_1$ as input and outputs $Y_1Y_2$, regarding $U_2$ as random. The better channel takes $U_2$ as input and outputs $U_1Y_1Y_2$. }
\end{figure}

\prlsection{Classical Polar Coding}Polar coding is based on the following simple construction. 
Let $W$ be a channel with binary input described by a random variable $X$ and output described by an arbitrary random variable $Y$. Now consider two instances of $W$, denoted $W_1$ and $W_2$,  whose inputs are connected by a \cnot gate, as shown in Fig.~\ref{fig:channeltransform}.

\vspace{-2mm}

\noindent For $U_1$, $U_2$ uniformly and independently distributed on $\{0,1\}$, it follows that
\begin{align}
2I(X:Y)&=I(X_1X_2:Y_1Y_2)=I(U_1U_2:Y_1Y_2)\nonumber\\
&=I(U_1:Y_1Y_2)+I(U_2:U_1Y_1Y_2),\label{eq:channelsplit}
\end{align}
since then $X_1$ and $X_2$ are uncorrelated; in the second line we have used the  chain rule for mutual information and the fact that $U_1$ and $U_2$ are independent. But $I(U_2:U_1Y_1Y_2)\geq I(X:Y)$, as $U_2=X_2$. 
We may then think of the \cnot gate as transforming the two physical channels into two logical channels corresponding to the two terms in~\eqref{eq:channelsplit}, whose input-output mutual informations are higher and lower than that of the original channel $W$, respectively. The ``better'' channel, denoted by $W_+$, has input $U_2$ and output $U_1Y_1Y_2$, while the ``worse'' channel $W_-$ has input $U_1$ and output $Y_1Y_2$.

This process can be recursively applied to $n=2^k$ instances of the channel $W$, resulting in a sequence of logical channels corresponding to all possible sequences of better and worse combinations of the channels at the previous stages. 
The original channels $W$ are first divided into two sets and the channel transform applied to pairs  of $W$s, one from each set. This produces $n/2$ channels apiece of types $W_+$ and $W_-$. Applying this procedure again to the channels of each type separately results in $n/4$ channels of each of the four types $W_{--}=(W_-)_-$, $W_{-+}$, $W_{+-}$, and $W_{++}$, and so on.  

\begin{figure}[h]
\centering

\beginpgfgraphicnamed{example4}
\def\vgap{.55}
\def\hgap{1.1}
\def\hgapfudge{.25}
\begin{tikzpicture}[scale=1]
\tikzstyle{empty} = [inner sep=1pt,outer sep=1pt]
\tikzstyle{gate} = [fill=white, draw]
\tikzstyle{ctrl} = [fill,shape=circle,minimum size=3pt,inner sep=0pt,outer sep=0pt]
\tikzstyle{targ} = [draw,shape=circle,minimum size=8pt,inner sep=0pt,outer sep=0pt]

\node[anchor=east] at (-3*\hgap,3*\vgap) (u0) {$U_1$};
\node[anchor=east] at (-3*\hgap,2*\vgap) (u1) {$U_2$};
\node[anchor=east] at (-3*\hgap,\vgap) (u2) {$U_3$};
\node[anchor=east] at (-3*\hgap,0) (u3) {$U_4$};

\node[anchor=west] at (1*\hgap,3*\vgap) (y0) {$Y_1$};
\node[anchor=west] at (1*\hgap,2*\vgap) (y1) {$Y_2$};
\node[anchor=west] at (1*\hgap,\vgap) (y2) {$Y_3$};
\node[anchor=west] at (1*\hgap,0) (y3) {$Y_4$};

\draw (u0) -- (y0);
\draw (u1) -- (y1);
\draw (u2) -- (y2);
\draw (u3) -- (y3);

\node[ctrl] at (-1*\hgap,\vgap) (ec) {};
\node[targ] at (-1*\hgap,3*\vgap) (et) {};
\draw (ec) -- (et.north);

\node[ctrl] at (-1.5*\hgap,0*\vgap) (ec) {};
\node[targ] at (-1.5*\hgap,2*\vgap) (et) {};
\draw (ec) -- (et.north);

\node[ctrl] at (-2.325*\hgap,2*\vgap) (g1c) {};
\node[targ] at (-2.325*\hgap,3*\vgap) (g1t) {};
\draw (g1c) -- (g1t.north);

\node[ctrl] at (-2.325*\hgap,0*\vgap) (g2c) {};
\node[targ] at (-2.325*\hgap,1*\vgap) (g2t) {};
\draw (g2c) -- (g2t.north);

\node[gate] at (0*\hgap,3*\vgap) (w0) {$W$};
\node[gate] at (0*\hgap,2*\vgap) (w1) {$W$};
\node[gate] at (0*\hgap,\vgap) (w2) {$W$};
\node[gate] at (0*\hgap,0) (w3) {$W$};

\end{tikzpicture}
\endpgfgraphicnamed
\caption{\label{fig:4chan} Recursive construction of the channel transformation for blocklength $n=4$. The physical channels $W$ are divided into two groups, the first $\sfrac n2$ and the last $\sfrac n2$, and the basic transformation of Fig.~\ref{fig:channeltransform} applied to pairs with one channel from each group. This results in $\sfrac n2$ channels each of $W_+$ and $W_-$; these channels are likewise each divided into two groups and the basic transformation applied again.}
\end{figure}
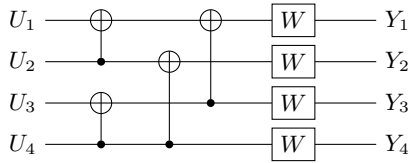 
\vspace{-3mm}
By appropriately grouping the channels, as shown in Figure~\ref{fig:4chan} for $n=4$, we can ensure that the $j$th logical channel has $U_j$ as its input and $\tilde{Y}_j := U_1\cdots U_{j-1}Y_1\cdots Y_n$ as its output. One can easily work out that $\frac n2\log_2 n$ \cnot gates are needed to implement the channel transform for blocklength $n=2^k$. Its  action is compactly described by the matrix $G_k = G^{\otimes k}$,
for $G = \bigl(\begin{smallmatrix} 1& 1\\ 0& 1
\end{smallmatrix} \bigr)$ over $\mathbbm{F}_2$.

The quality of the $j$th logical channel can be assessed by the similarity of the two possible output distributions $\tilde{Y}_{j}|U_j=0$ and $\tilde{Y}_{j}|U_j=1$, as measured by the fidelity 
\begin{align}
	F_j=\sum_{\tilde{y}}\sqrt{{\rm Pr}[\tilde{Y}_{j}= \tilde{y}|U_j=0]{\rm Pr}[\tilde{Y}_{j}=\tilde{y}|U_j=1]}.
\end{align}
 Finding the ``good'' logical channels which have an output fidelity below a predefined threshold can be done in $O(n)$ steps~\cite{mori_performance_2009}.  
For $n\rightarrow\infty$, the logical channels \emph{polarize}, their outputs becoming either identical or completely distinguishable, corresponding to useless or perfect channels, respectively. Moreover, the fraction of essentially perfect channels tends to the mutual information $I(X:Y)$,  the symmetric channel capacity~\cite{arikan_channel_2009}. 

To transmit information the encoder merely needs to use the good channels and fix or ``freeze'' the inputs to bad channels, making the inputs known to the decoder in advance~\footnote{Which values the inputs should take must also be determined. For a symmetric channel such as the BSC, all choices are equivalent, but for general channels they are not. In practice one could choose the inputs at random and be assured that the coding scheme will work with high probability.}.
All inputs $U_1\cdots U_{n}$ can then be decoded sequentially using maximum likelihood decoding. To determine $U_j$, the decoder decides for the input to the $j$th logical channel with the highest likelihood, since that channel's outputs $U_1\cdots U_{j-1}$ and $Y_1\cdots Y_{n}$ are available. If the $j$th input is frozen, then no decoding is necessary. The ratio of likelihoods for the two inputs inherits a recursive structure from the channel transformation, and using this it is possible to compute all the needed ratios using only $O(n\log n)$ operations~\cite{arikan_channel_2009}.

\prlsection{Quantum Channel Transformation}The central insight of this work is that the same channel transformation can be used to transmit quantum information. Let us regard the transformation as a unitary operator $V$ by fixing an orthonormal basis of $n$ qubits, the \emph{amplitude} basis $\{\ket{z}\}_{z\in\{0,1\}^n}$, and setting $V=\sum_{z\in\{0,1\}^n}\ketbra{G_kz}{z}$. In the complementary \emph{phase} basis, whose elements are given by $\ket{\wt{x}}=\frac{1}{\sqrt{2^n}}\sum_{z\in\{0,1\}^n}(-1)^{x\cdot z}\ket{z}$, $V$ acts as $G_k^T$: 
\begin{align}
V&=\tfrac{1}{2^n}\hspace{-5mm}\sum_{xx'z\in\{0,1\}^n}\hspace{-4mm}(-1)^{x'\cdot G_kz+x\cdot z}\ketbra{\wt{x}'}{\wt{x}}=\hspace{-2.5mm}\sum_{x\in\{0,1\}}\hspace{-2mm}\ketbra{\wt{G_k^Tx}}{\wt{x}}.
\end{align}
Here we have used $G_k^{-1}=G_k$. 
We could have anticipated this property from the fact that in the phase basis the \cnot gates act with control and target interchanged, as described by $G^T$. Since $G_k^T=(G^T)^{\otimes k}$, the action of $V$ in the phase basis is the same as in the amplitude basis, but with  inputs and outputs arranged in reverse order.

The dual behavior of $V$ can be used to construct quantum polar codes from classical polar codes. First let $W$ be a Pauli channel, which applies the operator $\sigma_x^u\sigma_z^v$ to the input with probability $p_{u,v}$, where $u,v\in\{0,1\}$ and $\sigma_x$ ($\sigma_z$) is the Pauli $x$ ($z$) operator.  Suppose that to $V$ and subsequently $W^{\otimes n}$ we input halves of maximally-entangled qubit pairs, each pair in the Bell state $\ket{\Phi}=\tfrac1{\sqrt{2}}\sum_{z\in\{0,1\}}\ket{z,z}$. Let $B$ denote the input qubits and $A$ the other halves of the pairs. Describing the channel as a unitary operation on $B$, also involving an auxiliary system $E$, this procedure results in the state 
\begin{align}
\label{eq:outputstate}
\ket{\Psi}^{ABE}=\tfrac1{\sqrt{2^n}}\!\!\!\!\!\!\!\!\!\!\sum_{u,v,z\in\{0,1\}^n}\!\!\!\!\!\!\!\!\!\sqrt{p_{u,v}^n}\ket{z}^A\sigma_z^v\sigma_x^u\ket{G_kz}^B\ket{u,v}^E\!\!,
\end{align}
where $\sigma_x^u$ denotes the operator $\sigma_x^{u_1}\otimes\cdots\otimes \sigma_x^{u_n}$ and $p_{uv}^n$ the probability distribution $p_{u_1,v_1}\cdots p_{u_n,v_n}$.
 
Now observe that an amplitude basis measurement of the $n$ $A$ systems with outcome $z$ leaves system $B$ in the amplitude basis state corresponding to $G_kz+u$, and that each outcome $z$ occurs with equal probability $\frac1{2^n}$. This is precisely the output one obtains when polar coding for a binary symmetric channel (BSC) with  bit flip probability $\delta_u=\sum_{v=0}^1p_{1,v}$. Let us call this channel the induced amplitude channel $W_A$. It is easy to work out that finding the phase of $A$ to be $x$ leaves $B$ in the phase basis state corresponding to $G_k^Tx+v$, also with uniform probability. This again corresponds to a BSC, the induced phase channel $W_P$, with  bit flip probability $\delta_v=\sum_{u=0}^1p_{u,1}$. 

The logical channels associated with the two induced channels both polarize, since each is essentially classical. Now suppose the $j$th input is good for both bases, which means that by using the logical channel outputs, the decoder can recreate the results of measuring the $A_j$ system in either the amplitude or phase basis (though not simultaneously, of course) and, as we will in the next section, this can be done efficiently. According to~\cite{renes_physical_2008}, the ability to determine both amplitude or phase implies that $A_j$ is maximally-entangled with the channel outputs. Moreover, the two classical decoding measurements used to recreate the amplitude or phase of $A_j$ can be combined to create an operation which creates the Bell pair $\ket{\Phi}$. Thus, to transmit quantum information, we simply need to make use of the inputs good for both amplitude and phase channels, and somehow freeze the remainder.  

Note that the outputs of the two logical channels are different; the  output of the $j$th amplitude channel is the collection $Z_1\dots Z_{j-1}B$ while the corresponding phase channel output is $X_{j+1}\dots X_{n}B$, where $Z_i$ ($X_i$) denotes the outcome of an amplitude (phase) basis measurement on $A_i$ from the state $\ket{\Psi}$ in \eqref{eq:outputstate}. Thus, the transformation $V$ does not cause quantum channel polarization \emph{per se}, as we are not dealing with a single channel. Rather, the two essentially classical channels polarize, and the fact that both amplitude and phase information are available to the decoder allows for quantum communication. 

\prlsection{Efficient Decoding}Now let us define the encoding scheme more precisely and show how the classical polar decoders can decode quantum inputs. To achieve the symmetric coherent information, we cannot simply combine the classical polar schemes of $W_A$ and $W_P$ as heuristically described above. Doing  so would ignore correlations between amplitude and phase errors, which are useful in the decoding process. Instead, we will make use of an extended version $W_{P'}$ of the phase channel, which takes $x$ to the pair $(x+v,u)$ with probability $p_{u,v}$. Note that this channel is also subject to classical channel polarization. 

Quantum information is encoded into inputs corresponding to good logical channels for both the amplitude and extended phase channels $W_A$ and $W_{P'}$. Call this set of inputs \mcQ. The remaining inputs fall into three subets: those  corresponding to logical channels bad for $W_A$ ($\mathcal{A}$), bad for $W_{P'}$ ($\mathcal{P}$), or bad for both ($\mathcal{E}$). Inputs to \mcA and \mcP are frozen in the amplitude and phase bases, respectively. The inputs to \mcE must be entangled with the decoder to ensure proper decoding; in a certain sense this allows the decoder to freeze the input in {both} bases. Thus, the code resulting from this construction is generically \emph{entanglement-assisted}. 

\begin{figure}[t]
\centering
\beginpgfgraphicnamed{decoder}
\def\dx{.67}
\def\dy{.75}
\def\gap{.1}
\begin{tikzpicture}[decoration={brace,amplitude=5}]
\tikzstyle{empty} = [inner sep=1pt,outer sep=1pt]
\tikzstyle{gate} = [fill=white, draw]
\tikzstyle{ctrl} = [fill,shape=circle,minimum size=3pt,inner sep=0pt,outer sep=0pt]
\tikzstyle{targ} = [draw,shape=circle,minimum size=8pt,inner sep=0pt,outer sep=0pt]

\node[anchor=east] at (-.8*\dx,3*\dy) (A) {$A_\mcE$};
\node[anchor=east] at (-.8*\dx,2*\dy) (B) {$B$};
\node[anchor=east] at (-.8*\dx,1*\dy) (C) {$C$};
\node[anchor=east] at (-.8*\dx,0*\dy) (D) {$D$};

\node[anchor=east] at (0.,1*\dy) (Ck) {$\ket{0}$};
\node[anchor=east] at (0.,0*\dy) (Dk) {$\ket{0}$};

\node[anchor=west] at (10.75*\dx,3*\dy) (Ar) {};
\node[anchor=west] at (10.75*\dx,2*\dy) (Br) {};
\node[anchor=east] at (10.75*\dx,1*\dy) (Cr) {$\ket{u}$};
\node[anchor=east] at (10.75*\dx,0*\dy) (Dr) {$\ket{\wt{v}}$};

\draw (A) -- (Ar);
\draw (B) -- (Br);
\draw (Ck) -- (Cr);
\draw (Dk) -- (Dr);


\node[ctrl] at (1*\dx,3*\dy) (u1e) {};
\node[targ] at (1*\dx,1*\dy) (u1t) {};
\draw (u1e) -- (u1t.south);
\node[gate,minimum height=20pt] at (1*\dx,2*\dy) (u1m) {$\mcM_A$};

\node[gate] at (2*\dx,1*\dy) (u2v) {$V$};

\node[ctrl] at (3*\dx,2*\dy) (u3c) {};
\node[targ] at (3*\dx,1*\dy) (u3t) {};
\draw (u3c) -- (u3t.south);

\node[ctrl] at (4*\dx-2*\gap,1*\dy) (u4c) {};
\node[targ] at (4*\dx-2*\gap,2*\dy) (u4t) {};
\draw (u4c) -- (u4t.north);

\draw [decorate] (4*\dx,-0.5*\dy) -- (0.25*\dx,-0.5*\dy);
\node [anchor=north] at (2.125*\dx,-.75*\dy) {\footnotesize Amplitude recovery};

\node[gate] at (5*\dx-.5*\gap,3*\dy) (h1) {$H$};

\node[ctrl] at (6*\dx-2*\gap,3*\dy) (v1e) {};
\node[targ] at (6*\dx-2*\gap,0*\dy) (v1t) {};
\draw (v1e) -- (v1t.south);
\node[gate,minimum height=32pt] at (6*\dx+-2*\gap,1.5*\dy) (u1m) {$\mcM_{P'}$};

\node[gate] at (7*\dx-3.5*\gap,3*\dy) (h2) {$H$};
\node[gate] at (7*\dx-2*\gap,0*\dy) (h3) {$H$};

\node[gate] at (8*\dx-1*\gap,0*\dy) (vv) {$V$};

\node[ctrl] at (9*\dx-3*\gap,0*\dy) (v3c) {};
\node[targ] at (9*\dx-3*\gap,2*\dy) (v3t) {};
\draw (v3c) -- (v3t.north);


\node[ctrl] at (10*\dx-5.5*\gap,2*\dy) (vcc) {};
\node[targ] at (10*\dx-5.5*\gap,0*\dy) (vct) {};
\draw (vcc) -- (vct.south);

\draw [decorate] (9.5*\dx,-0.5*\dy) -- (4.5*\dx,-0.5*\dy);
\node [anchor=north] at (7*\dx,-.75*\dy) {\footnotesize Phase recovery};

\node[gate] at (9.85*\dx+\gap,2*\dy) (cne) {$V$};

\end{tikzpicture}
\endpgfgraphicnamed
\caption{\label{fig:dec}The quantum decoding circuit. With the help of the two ancilla systems $C$ and $D$, the decoding measurements $\mcM_A$ and $\mcM_{P'}$ for $W_A$ and $W_{P'}$ are used to diagnose and correct the amplitude and phase error patterns $u$ and $v$, respectively. The final step reverses the original encoding.}
\vspace{-2mm}
\end{figure}
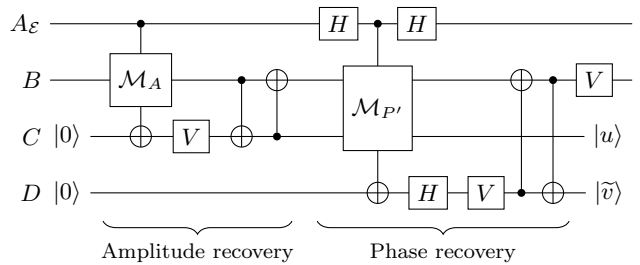

The quantum decoder is constructed from the classical decoders of $W_A$ and $W_{P'}$. In the language of quantum theory, these can be regarded as generalized measurements $\mcM_A$ and $\mcM_{P'}$, respectively, for each is an operation on $B$ (dependent on the frozen bits) having a classical output: a guess of the input of the corresponding channel. The basic idea of the quantum decoder, shown in Figure~\ref{fig:dec}, is to coherently run the two classical decoders in succession, determining and correcting the amplitude and phase error patterns $u$ and $v$. 

To see that it works as intended, suppose the sender encodes halves of $\ket{\Phi}$ into \mcQ, amplitude basis states corresponding to the classical bitstring $g$ into \mcA, and phase basis states corresponding to the bitstring $h$ into \mcP. The \mcE inputs are also halves of $\ket{\Phi}$, but with the other half held by the receiver. The quantum state describing the systems after encoding and transmission can be expressed as
$\ket{\Psi_1}^{ABE}={\mathcal{N}}\Pi^{\mcA}_{g}\widetilde{\Pi}^{\mcP}_{h}\ket{\Psi}^{ABE}$, where $\Pi^{\mcA}_g$ is the projector onto the string $g$ in the amplitude basis of the systems $A$ in the set \mcA, and similarly for $\widetilde{\Pi}^\mcP$ in the phase basis, while $\mathcal{N}$ is the normalization factor $\sqrt{|\mcA\cup\mcP|}$. 

The first step in the quantum decoder is to use the classical $W_A$ decoder to determine the amplitude error pattern $u$ and correct it. This requires the amplitude-frozen input to \mcA and \mcE, of which the former, $g$, is known, and the latter, call it $g'$ can be generated by measuring $A_\mcE$ in the amplitude basis. Together $g$ and $g'$ comprise the frozen bits $f_A$ needed by the classical decoder to determine the amplitude input $z$. The quantum decoder performs the $A_\mcE$ measurement coherently, i.e.\ controlling the measurement operation on $B$ as shown in Fig.~\ref{fig:dec}, and stores the result in an ancillary system $C$. If the classical decoder has a low error probability, then the state $\ket{\Psi_2}$ resulting from this process is essentially equal to 
\begin{align*}
\ket{\Psi_2}\approx \sum_{u,v,z}\sqrt{p^n_{uv}}\Pi^{\mcA}_{g}\widetilde{\Pi}^{\mcP}_{h}\ket{z}^A\sigma_z^v\ket{G_kz+u}^B\ket{z}^C\ket{u,v}^E,
\end{align*}
where we have disregarded normalization for simplicity. Observe that the channel output $B$ and entanglement assistance system $A_\mcE$ are essentially unchanged in this process; in the Appendix we give the precise details of the approximation.
The error pattern $u$ can  be transferred to $C$ by first applying the encoding circuit $V$ adding the amplitude value of $B$ to the result with a \textsc{cnot} operation. Amplitude errors can then be corrected by another \cnot operation. This results in the state
\begin{align*}
\ket{\Psi_3}
&\approx\sum_{u,v,x}\sqrt{p^n_{uv}}\Pi^{\mcA}_{g}\widetilde{\Pi}^{\mcP}_{h}\ket{\wt{x}}^A\ket{{G_k^T\wt{x}+\wt{v}}}^B\ket{u}^C\ket{u,v}^E,
\end{align*}
where we have abused notation in writing $G_k^T\wt{x}+\wt{v}$.

Proceeding analogously, the phase error pattern $v$ can be diagnosed and corrected by using the 
$W_{P'}$ decoder on systems $BC$. Working in the phase basis requires a few extra Hadamard gates and inverts the \cnot gates in the final two operations. After correcting the phase errors, the encoding operation can be reversed, leaving the state
 \begin{align*}
 \ket{\Psi_4}&\approx \sum_{u,v,x}\sqrt{p^n_{uv}}\Pi^{\mcA}_{g}\widetilde{\Pi}^{\mcP}_{h}\ket{\wt{x}}^A\ket{\wt{x}}^B\ket{u}^C\ket{\wt{v}}^D\ket{u,v}^E,
\end{align*}
which describes a maximally-entangled state between the \mcQ systems in $A$ and $B$.

This protocol achieves a net rate given by $R=\lim_{n\rightarrow \infty}\frac1n[\log|\mcQ|-\log|\mcE|]$. By the properties of classical polar coding, for any $\epsilon>0$ we can choose $n$ large enough such that $\log|\mcA\cup \mcE|\leq n[1-I(W_A)+\epsilon]$ and $\log|\mcP\cup\mcE|\leq n[1-I(W_{P'})+\epsilon]$, where $I(W_A)$ denotes the input-output mutual information of the channel $W_A$ with uniform inputs, and similarly for $I(W_{P'})$. Since \mcQ, \mcA, \mcP, and \mcE are disjoint, this yields $R\geq I(W_A)+I(W_{P'})-1$. Direct calculation gives $R\geq 1-H(p_{uv})$, which is precisely $-H(A|B)$ for the state $\ket{\Phi}^{AB}$ after subjecting $B$ to the Pauli channel $W$. Here, $H$ denotes the Shannon entropy of a classical distribution or von Neumann entropy of a quantum state. Thus, the protocol achieves the coherent information for $W$ with input $\ket{\Phi}$.

\prlsection{Coding Without Entanglement Assistance}Although entanglement assistance is used in the above construction, nowhere is it shown to be necessary. We now give a condition under which the rate of entanglement assistance required is zero, namely,
\begin{align}
F_{A,0}+F_{P',0}\leq 1,
\end{align}
where $F_{A,0}$ ($F_{P',0}$) is the output fidelity of the amplitude (extended phase) channel. To show this result, we follow the original technique of~\cite{arikan_channel_2009,arikan_rate_2008,arikan_rate_2009} establishing the channel polarization phenomenon. 

Consider the stochastic process consisting of random variables $C_j\in\{0,1\}$ for $j=0,\dots,n$, corresponding to choices of better ($C_j=1$) and worse ($C_j=0$) channels in the recursive channel construction, with equal probability for each. The entire sequence $C_0,\dots,C_n$ corresponds to an input to the polar coding circuit. Closely related is the process describing the output fidelity $F_j$ of the channel corresponding to the sequence $C_0,\dots,C_j$. The idea behind the polarization proofs in~\cite{arikan_channel_2009,arikan_rate_2009} is to show that in the limit $n\rightarrow \infty$, $F_n$ converges to a random variable having support solely on $\{0,1\}$, meaning the input to a particular channel is either transmitted perfectly ($F_n\rightarrow 0$) or completely garbled ($F_n\rightarrow 1$).  

One method of examining the convergence of $F_n$ is to bound it by another process whose convergence properties are easier to determine. To this end, \arikan{} and Telatar considered the  process $F_j'$~\cite{arikan_rate_2008,arikan_rate_2009}, defined by 
\begin{align}
F'_{j+1}=\left\{\begin{array}{ll}{F'_k}^{2}& \quad C_j=0\\
2F_j'-{F'_j}^{2}& \quad C_j=1
\end{array}\right.,
\end{align}
with $F'_0=F_0$, for which it can be shown that $F_j\leq F'_j$ for
all $j$.
The $F_j'$ process has the property that for each sequence $C_j$ there exists a threshold initial value $F'_{\rm th}$  below which $\lim_{n\rightarrow \infty}F'_{n}=0$ and above which $\lim_{n\rightarrow \infty}F'_{n}=1$ (see Observation 4 of~\cite{arikan_rate_2008}). Additionally, observe that the new process is invariant under the map taking $F_j'$ to $1-F_j'$ and $C_j$ to $1-C_j$.

In the quantum case we are interested in the fidelity processes of the amplitude and phase channels, denoted by $F_{A,j}$ and $F_{P',j}$, respectively. The fact that the phase encoder is the reverse of the amplitude encoder implies that $F_{P',j}$ makes the opposite channel choice as $F_{A,n}$ at each step, i.e.\ $C_{P',j}=1-C_{A,j}$. Considering the associated processes $F_{A,n}'$ and $F_{P',n}'$ and their symmetries as described above, we therefore find that the sum process $F_{A,n}'+F_{P',n}'$ only converges to 2 when $F_{A,0}'+F_{P',0}'\geq F'_{\rm th}+1-F'_{\rm th}=1$. 
But because $F'_{A,0}=F_{A,0}$ and $F_{P',0}'=F_{P',0}$ and $F'_j\geq F_j$, it follows that $F_{A,n}+F_{P',n}$ can only converge to 2 when $F_{A,0}+F_{P',0}\geq 1$ (though it might still converge to 1 or 0). A value of 2 for the sum process indicates that the input  is useless for transmitting both amplitude and phase information, and therefore we conclude that $F_{A,0}+F_{P',0}\leq 1$ implies $\mcE=\emptyset$.

We can easily reformulate this threshold result in terms of noise rates of two Pauli channels of interest: independent amplitude and phase errors, and the depolarizing channel. In the former case $W_P$ is as good as $W_{P'}$ since the error patterns are independent; both $W_A$ and $W_P$ are binary symmetric channels for which the fidelity is given by $2\sqrt{\delta(1-\delta)}$ for $\delta$ the bit flip probability. Thus, the above condition becomes  $2\sqrt{\delta_u(1-\delta_u)}+2\sqrt{\delta_v(1-\delta_v)}\leq 1$. In the case of equal error rates, we find a threshold of $(2-\sqrt{3})/4\approx 6.70\%$. For comparison, the coherent information of that channel goes to zero at 11.00\%.
For the depolarizing channel with parameter $q$, a full calculation using $W_{P'}$ leads to the condition $2\sqrt{\frac{2q}{3}(1-\frac{2q}{3})}+\frac{2q}{3}+2\sqrt{(1-q)\frac q3}\leq 1$. This yields a threshold of approximately 12.05\%, compared with the coherent information threshold of approximately 18.93\%.

\prlsection{Conclusions}We have adapted the results for classical polar codes to show that there likewise exist efficiently encodable and decodable qubit codes, which when entanglement-assisted, achieve a communication rate equal to the coherent information for Pauli channels. Our construction also applies to the quantum erasure channel with erasure probability $p$, as the outputs of the associated classical channels $W_A$ and $W_P$ are again classical (simultaneously diagonalizable). In fact, as $W_A$ and $W_P$ are also erasure channels, the quantum polar coding scheme achieves the capacity of the erasure channel, namely $1-2p$~\cite{bennett_capacities_1997}. It is also easy to see from the condition presented in the previous section that, in this case, the rate of entanglement assistance required is zero.

An immediate practical application of such codes which is feasible using current technology is to quantum key distribution (QKD). Due to the CSS nature of the codes, the well-known relationship between CSS coding and secret key generation~\cite{shor_simple_2000} implies that our protocol can be converted into a means for efficient, high-rate secret key distillation, possibly assisted by a preshared classical key. Such a key distillation scheme would be suitable for use in prepare-and-measure QKD protocols, as the polar code-based key distillation step itself would be entirely classical and not require actually implementing the full quantum code using a quantum computer. Quantum polar codes may also prove useful in the study of fault-tolerant quantum computation as an alternative to concatenation-based approaches.  

Our results merely initiate the study of quantum polar codes, and many unanswered questions remain. Most immediate is the issue of entanglement assistance. We have been able to rigorously show that under low-noise conditions, the rate of entanglement assistance required is zero, but one would like to know that this is always the case, a conjecture supported by preliminary numerical evidence. 

Together with Wilde, one of us has shown how to combine the method here with~\cite{wilde_polar_2011-1} to construct a polar coding scheme for arbitrary qubit channels and show that the rate at which entanglement assistance is required goes to zero for degradable channels~\cite{wilde_quantum_2012}. One would like to extend the method to general qudit channels, as well as investigate how the normal, not necessarily symmetric, coherent information can be achieved.

\noindent {\bf Acknowledgements.} We thank Omar Fawzi and Mark M.\ Wilde for helpful conversations. This work was supported by the Swiss National Science Foundation (SNF) through the National Centre of Competence in Research ``Quantum Science and Technology'' and project No. 200020-135048, as well as by the European Research Council (ERC), via grant No. 258932. FD acknowledges support of the Natural Science and Engineering Research Council (NSERC) of Canada via the postdoctoral fellowship program. 

\bibliography{polar}


\clearpage

\widetext
\appendix
\section*{Appendix}

This appendix demonstrates that the quantum polar decoder for channel $W$ outputs high-fidelity entanglement when the classical polar decoders of the associated classical channels $W_A$ and $W_{P'}$ have low error probability. 

In the classical polar coding scheme for the amplitude channel $W_A$, the frozen bits $f$ and message bits $m$ are sent through the encoder and channel to the receiver. The frozen bits are input to the set $\mcA\cup\mcE$ while the message bits are input to $\mcQ\cup\mcP$. We denote by $z$ the input to the polar coding circuit; to emphasize the dependence of this string on $m$ and $f$, we write $z(m,f)$. In the notation of quantum information theory, the output of the channel is the mixed state 
\begin{align}
\rho_{m;f}^B=\sum_u p_u^n \ketbra{G_kz(m,f)+u}{G_kz(m,f)+u}^B.
\end{align} 
The receiver attempts to recover $m$ from the channel output $\rho_{m;f}^B$ using the polar decoding algorithm and the frozen bits $f$. As discussed in the main text, this process can be viewed as a measurement $\mcM_A$ of $\rho_{m;f}$. This measurement consists of elements $\Lambda_{f;z}^B$ which determine the probability that the decoder guesses that the entire input was $z$ via the expression ${\rm Pr}(z|m,f)={\rm Tr}[\rho_{m;f}^B\Lambda^B_{f,z}]$. This is equivalent to the probability of guessing the input message was $m$ since the frozen bits are known to the decoder with certainty. 
Thus, the probability of incorrect decoding averaged over messages $m$ and frozen bits $f$ is given by
\begin{align}
p_{{\rm err}}(\mcM_A)=1-\frac{1}{2^n}\sum_{m\in\{0,1\}^{\log|\mcQ\cup\mcP|},f}{\rm Tr}[\rho_{m;f}^B\Lambda_{f,z(m,f)}^B].
\end{align}

The situation for the extended phase channel $W_{P'}$ can be expressed similarly. Here the message bits $m$ are input to the set $\mcQ\cup\mcA$ while the frozen bits $f$ are input to $\mcP\cup\mcE$, and we call the entire input to the polar coding circuit $x$. Assuming the channel acts on its input in the phase basis (which will be convenient later), the output state can be written 
\begin{align}
\sigma_{m;f}^{BC}=\sum_{u,v}p_{u,v}^n\ketbra{\wt{G_k^Tx(m,f)+v}}{\wt{G_k^Tx(m,f)+v}}\otimes\ketbra{u}{u}^{C}.
\end{align}
The measurement $\mcM_{P'}$ has elements $\Gamma_{h,x}^{BC}$ such that the average error probability is 
\begin{align}
p_{{\rm err}}(\mcM_{P'})=1-\frac{1}{|\mcQ\cup\mcA|}\sum_{m\in\{0,1\}^{\log|\mcQ\cup\mcA|},f}{\rm Tr}[\sigma_{m;f}^{BC}\Lambda_{f,x(m,f)}^{BC}].
\end{align}

Now consider the quantum state $\ket{\Psi_1}^{ABE}$ from the main text, describing the input and output systems in the quantum polar coding scheme, with amplitude input $g$ in \mcA and $h$ in \mcP. It takes the form
\begin{align}
\ket{\Psi_1}^{ABE}=\sqrt{\frac{|\mcA\cup\mcP|}{2^n}}\sum_{u,v,z\in\{0,1\}^n}\!\sqrt{p_{u,v}^n}\,\Pi_{g}^\mcA\wt{\Pi}_h^\mcP\ket{z}^A(-1)^{v\cdot (G_kz+u)}\ket{G_kz+u}^B\ket{u,v}^E\!\!,
\end{align}
where we have explicitly written out the action of $\sigma_z^v$. 
The first step of the decoder is to coherently implement the $\mcM_A$ measurement. This is accomplished by coherently measuring the $\mcE$ subsystem of $A$ to determine $g'$, and thus the entire string of frozen bits $f$,  subsequently using $f$ to coherently implement $\Lambda_{f,z}^B$, and storing the result $z$ in an ancillary system $C$. Formally, the state resulting from this transformation can be expressed as 
\begin{align}
\ket{\Psi_2}^{ABCE}=\sqrt{\frac{|\mcA\cup\mcP|}{2^n}}\sum_{u,v,z,z'\in\{0,1\}^n,g'\in\{0,1\}^{\log|\mcE|}}\!\!\!\!\!\!\!\!\!\sqrt{p_{u,v}^n}\,\Pi_{g'}^\mcE\Pi_{g}^\mcA\wt{\Pi}_h^\mcP\ket{z}^A(-1)^{v\cdot (G_kz+u)}\sqrt{\Lambda_{f,z'}}^B\ket{G_kz+u}^B\ket{z'}^C\ket{u,v}^E\!\!.
\end{align}
Ideally, the output would be the state 
\begin{align}
\ket{\Psi_2'}^{ABCE}=\sqrt{\frac{|\mcA\cup\mcP|}{2^n}}\sum_{u,v,z\in\{0,1\}^n}\!\!\sqrt{p_{u,v}^n}\,\Pi_{g}^\mcA\wt{\Pi}_h^\mcP\ket{z}^A(-1)^{v\cdot (G_kz+u)}\ket{G_kz+u}^B\ket{z}^C\ket{u,v}^E\!\!,
\end{align}
i.e.\ the value of $z$ in $A$ would simply be copied to $C$ without any backaction on systems $B$ and $A_\mcE$ at all. Computing the fidelity of the two states and averaging over uniformly-random choices of $g$, we find
\begin{align}
\frac{1}{|\mcA|}\sum_{g\in\{0,1\}^{\log|\mcA|}}\braket{\Psi_2|\Psi'_2}&=\frac{|\mcP|}{2^n}\sum_{g,g',u,v,z}p_{u,v}^n\bra{z}\Pi_{g'}^\mcE\Pi_g^\mcA\wt{\Pi}_h^\mcP\ket{z}\bra{G_kz+u}\sqrt{\Lambda_{f,z}}\ket{G_kz+u}\\
&=\frac{|\mcP|}{2^n}\sum_{u,v,m,f}p_{u,v}^n\bra{m}\wt{\Pi}_h^\mcP\ket{m}^{\mcQ\cup\mcP}\bra{G_kz(m,f)+u}\sqrt{\Lambda_{f,z(m,f)}}\ket{G_kz(m,f)+u}\\
&=\frac1{2^n}\sum_{m,f}{\rm Tr}[\rho_{m,f}\sqrt{\Lambda_{f,z(m,f)}}]\\
&\geq \frac{1}{2^n}\sum_{m,f}{\rm Tr}[\rho_{m,f}{\Lambda_{f,z(m,f)}}]\\
&=1-p_{\rm err}(\mcM_A).
\end{align}
In the second equality we relabel the sum on $z$ as a sum on $f'$ in $\mcA\cup\mcE$ and a sum on $m$ in $\mcQ\cup\mcP$; the projections on $g$ and $g'$ ensure $f'=f$ in the former summation. In the third equality we use the fact that the outcomes of a conjugate basis measurement ($\wt{\Pi}_h$) have uniform probability for an amplitude-basis state ($\ket{m}$). Then we are able to marginalize $p_{u,v}$ over $v$ and obtain the state $\rho_{m;f}$. Finally, the inequality arises because $\sqrt{\Lambda}\geq \Lambda$ for any operator $0\leq \Lambda\leq \mathbbm{1}$. 

Since the average of the fidelities for different choices of $g$ exceeds $1-p_{\rm err}(\mcM_A)$, there certainly exists one value of $g$ for which this is true, and we choose this one for the coding scheme. In fact, by Markov's inequality applied to $1-\braket{\Psi_2|\Psi_2'}$, a fraction at most  $\sqrt{p_{\rm err}(\mcM_A)}$ choices have fidelity lower than $1-\sqrt{p_{\rm err}(\mcM_A)}$.  Converting the fidelity bound into trace distance~\citesupp{nielsen_quantum_2000}, we find that the actual and ideal outputs have trace distance no greater than $\sqrt{2p_{\rm err}(\mcM_A)}$. Thus, we may proceed with the action of the decoder using the ideal output. 

The remainder of the amplitude recovery step produces the state 
\begin{align}
\ket{\Psi_3}^{ABCE}&=\sqrt{\frac{|\mcA\cup\mcP|}{2^n}}\sum_{u,v,z\in\{0,1\}^n}\!\!\sqrt{p_{u,v}^n}\,\Pi_{g}^\mcA\wt{\Pi}_h^\mcP\ket{z}^A\sigma_z^v\ket{G_kz}^B(-1)^{u\cdot v}\ket{u}^C\ket{u,v}^E\\
&=\sqrt{\frac{|\mcA\cup\mcP|}{2^n}}\sum_{u,v,x\in\{0,1\}^n}\!\!\sqrt{p_{u,v}^n}\,\Pi_{g}^\mcA\wt{\Pi}_h^\mcP\ket{\wt{x}}^A\ket{\wt{G_k^Tx+v}}^B(-1)^{u\cdot v}\ket{u}^C\ket{u,v}^E.
\end{align}
Measuring this state coherently with $\mcM_{P'}$ in the same manner as before produces
\begin{align}
\ket{\Psi_4}^{ABCDE}=\sqrt{\frac{|\mcA\cup\mcP|}{2^n}}\sum_{u,v,x,x'\in\{0,1\}^n,h'}\!\!\sqrt{p_{u,v}^n}\,\wt{\Pi}_{h'}^\mcE\Pi_{g}^\mcA\wt{\Pi}_h^\mcP\ket{\wt{x}}^A\sqrt{\Gamma_{f;x'}}^{BC}\ket{\wt{G_k^Tx+v}}^B(-1)^{u\cdot v}\ket{u}^C\ket{\wt{x}'}^D\ket{u,v}^E.
\end{align}
As before, the ideal output would see a copy of $\wt{x}$ from $A$ in system $D$, with no backaction on $A_\mcE$, $B$, or $C$:
\begin{align}
\ket{\Psi_4'}^{ABCDE}=\sqrt{\frac{|\mcA\cup\mcP|}{2^n}}\sum_{u,v,x\in\{0,1\}^n}\!\!\sqrt{p_{u,v}^n}\,\Pi_{g}^\mcA\wt{\Pi}_h^\mcP\ket{\wt{x}}^A\ket{\wt{G_k^Tx+v}}^B(-1)^{u\cdot v}\ket{u}^C\ket{\wt{x}}^D\ket{u,v}^E.
\end{align}
By an entirely similar calculation as in the amplitude case, we find that the error probability of $\mcM_{P'}$ bounds the averaged fidelity of the actual and ideal outputs:
\begin{align}
\frac1{|\mcP|}\sum_{h\in\{0,1\}^{\log|\mcP|}}\braket{\Psi_4|\Psi_4'}\geq 1-p_{\rm err}(\mcM_{P'}).
\end{align}
Thus we may proceed as before, choosing the optimal value of $h$ for the coding scheme and continuing the decoder analysis under the assumption that the output is ideal, and deal with the accumulated errors later. The remainder of the phase recovery and correction operation produces 
\begin{align}
\ket{\Psi_5}^{ABCDE}=\sqrt{\frac{|\mcA\cup\mcP|}{2^n}}\sum_{u,v,x\in\{0,1\}^n}\!\!\sqrt{p_{u,v}^n}\,\Pi_{g}^\mcA\wt{\Pi}_h^\mcP\ket{\wt{x}}^A\ket{\wt{G_k^Tx}}^B(-1)^{u\cdot v}\ket{u}^C\ket{\wt{v}}^D\ket{u,v}^E.
\end{align}
Finally, a further application of the encoding circuit to $B$ produces maximally entangled qubit pairs in the \mcQ and \mcE subsystems of $A$ and $B$. The latter systems are local to Bob, but the former systems represent shared entanglement between Alice and Bob. 

By using the triangle-inequality property of the trace distance, the actual state of $A_\mcQ$ and $B_\mcQ$ produced by the decoder has distance less than $\sqrt{2p_{\rm err}(\mcM_A)}+\sqrt{2p_{\rm err}(\mcM_{P'})}$ to the ideal, maximally entangled output. Thus, the quantum decoder functions well whenever the classical decoders for $W_A$ and $W_{P'}$ do, too.

\bibliographystylesupp{unsrt}
\bibliographysupp{polar}
\end{document}